\documentclass{article}

\usepackage{PRIMEarxiv}

\usepackage[utf8]{inputenc} 
\usepackage[T1]{fontenc}    
\usepackage{hyperref}  
\hypersetup{
    colorlinks=true,
    linkcolor=blue,
    filecolor=blue,      
    urlcolor=blue,
    citecolor=blue,
}     
\usepackage{url}            
\usepackage{booktabs}       
\usepackage{amsfonts} 
\usepackage{amsmath}
\usepackage{nccmath}
\usepackage{tabularx}
\usepackage{graphicx}
\usepackage{caption}  
\DeclareCaptionFormat{sanslabel}{#3}    
\usepackage{nicefrac}       
\usepackage{microtype}      
\usepackage{lipsum}
\usepackage{fancyhdr}       
\usepackage{graphicx} 
\usepackage{braket}      
\graphicspath{{media/}}     

\title{Machine Learning-Assisted Multi-Objective Optimization of Battery Manufacturing from Synthetic Data Generated by Physics-Based Simulations}

\author{ {\hspace{1mm}Marc Duquesnoy}$^{1, 2}$ \\
	\texttt{} \\
	\And
	{\hspace{1mm}Chaoyue Liu}$^{1, 3}$ \\
	\texttt{} \\
	\And
	{\hspace{1mm}Diana Zapata Dominguez}$^{1, 3}$ \\
	\texttt{} \\
	\And
	{\hspace{1mm}Vishank Kumar}$^{2, 4}$ \\
	\texttt{} \\
	\And
	{\hspace{1mm}Elixabete Ayerbe}$^{2, 5}$ \\
	\texttt{} \\
	\And
	{\hspace{1mm}Alejandro A. Franco}$^{1, 2, 3, 6, *}$}

\begin{document}
\maketitle

{$^{1}$ \small Laboratoire de Réactivité et Chimie des Solides (LRCS), Université de Picardie Jules Verne UMR CNRS 7314, Hub de l’Energie, 80039 Amiens, France.}

{$^{2}$ \small Alistore-ERI European Research Institute, CNRS FR 3104, Hub de l’Energie, 80039 Amiens, France.}

{$^{3}$ \small Réseau sur le Stockage Electrochimique de l’Energie (RS2E), FR CNRS 3459, Hub de l’Energie, 80039 Amiens, France.}

{$^{4}$ \small Umicore Group Research \& Development, Olen, Belgium.}

{$^{5}$ \small CIDETEC, Basque Research and Technology Alliance (BRTA), Pº Miramón 196, 20014 Donostia-San Sebastian, Spain.}

{$^{6}$ \small Institut Universitaire de France, 103 Boulevard Saint Michel, 75005 Paris, France.}

{$^{*}$ \small Corresponding author : alejandro.franco@u-picardie.fr (Alejandro A. Franco)}

\vspace{2cm}

\begin{abstract}
The optimization of the electrodes manufacturing process constitutes one of the most critical steps to ensure high-quality Lithium-Ion Battery (LIB) cells, in particular for automotive applications. Because LIB electrode manufacturing is a complex process involving multiple steps and interdependent parameters, we have shown in our previous works that 3D-resolved physics-based models constitute very useful tools to provide insights about the impact of the manufacturing process parameters on the textural and performance properties of the electrodes. However, their high-throughput application for electrode properties optimization and inverse design of manufacturing parameters is limited due to the high computational cost associated with this kind of model. In this work, we tackle this issue by proposing an innovative approach, supported by a deterministic machine learning (ML)-assisted pipeline for multi-objective optimization of LIB electrode properties and inverse design of its manufacturing process. Firstly, the pipeline generates a synthetic dataset from physics-based simulations with low discrepancy sequences, that allow to sufficiently represent the manufacturing parameters space. Secondly, the generated dataset is used to train deterministic ML models for the implementation of a fast multi-objective optimization, to identify an optimal electrode and the manufacturing parameters to adopt in order to fabricate it. Lastly, this electrode was successfully fabricated experimentally, proving that our modeling pipeline prediction is physical-relevant. Here, we demonstrate our pipeline for the simultaneous minimization of the electrode tortuosity factor and maximization of the effective electronic conductivity, the active surface area, and the density, all being parameters that affect the Li+ (de-)intercalation kinetics, ionic, and electronic transport properties of the electrode.
\end{abstract}

\keywords{Battery Manufacturing \and Bayesian Optimization \and Physics-based Modeling \and Sobol Sequences \and Machine Learning}

\section{Introduction}
\paragraph{} Lithium-Ion Batteries (LIBs) represents the leading technology in the ongoing energy transition \cite{1,2}. Such technology is deployed in many domains like portable devices and electric vehicles, due to their high performances and relatively good cell durability, while efforts leading to the emergence of gigafactories aim to decrease their massive production cost \cite{3,4}. The latest signs of progress in the spectacular increase of cycle life,\cite{5} higher energy and power densities,\cite{6} are not only linked to the type of materials adopted but also to a meticulous optimization of the battery cell manufacturing process \cite{7}. Nevertheless, such an optimization currently relies to trial-and-error approaches which are time consuming and costly. It is believed that manufacturing scrap rates are comprised between 5 and 30\,\%, the lowest values being more representative of production and the highest ones of prototyping activities \cite{8}. Indeed, the manufacturing process of LIBs is a complex process involving multiple interlinked steps and process parameters. Such steps are the electrode slurry preparation through the active material, carbon additive binder mixing in a solvent, the slurry coating and drying, the calendering of the resulting electrode, the cell assembly, the electrolyte filling, and the formation \cite{9}. Examples of the process parameters associated to these steps are the slurry formulation and solid content, the coating speed, the drying rate and the calendering pressure. Furthermore, trial-and-error optimization does not guarantee the achievement of the best electrodes, because multi-objective optimization (\textit{i.e.} maximizing/minimizing multiple properties at the same time) is very difficult if it is only based on empirical approaches supported for instance on design of experiments \cite{10}. There is also a lack of tools able to tell which process parameters one needs to adopt in order to manufacture an electrode with desired properties and, ultimately, the best electrode as possible. 
 
\paragraph{} That said, the digitalization of the manufacturing process of LIBs is called to bring powerful tools to advance in the understanding of how manufacturing parameters impact electrode and cell properties (\textit{e.g.} electrode porosity, tortuosity factor, conductivity, cell capacity), and to perform such optimization \cite{11,12,13,14}. This digitalization is expected to be supported on physics-based and machine learning (ML) modeling simulating each step of the manufacturing process and their interlinks. Our ERC-funded ARTISTIC project \cite{15} pioneered this, by bringing to the community a series of unique computational 3D-resolved models describing each step of the manufacturing process and being sequentially linked with each other, \textit{i.e.} the output of a model is the input of the following model, and so on. For instance, Coarse-Grained Molecular Dynamics (CGMD) simulations are used to simulate in 3D, electrode slurries and their drying resulting in synthetic uncalendared electrode microstructures (called mesostructures in the following) \cite{13,16,17}. While the Discrete Element Method (DEM) is used to simulate the calendering of such mesostructures \cite{13}. The Lattice Boltzmann Method (LBM) is used to simulate the electrolyte infiltration of the calendered electrodes taken alone or in the cell sandwich \cite{16}. In addition, we directly use the results from these models into 4D electrochemical models simulating galvanostatic discharge-charge and electrochemical impedance spectroscopy, unlocking the relationships between manufacturing parameters, electrode mesostructure and electrochemical performances \cite{18,19,20}. These models are carefully validated in our experimental battery prototyping platform. They are being integrated by us in an online calculator, usable through any Internet Browser, allowing to simulate the LIB manufacturing process without the need of computational skill. By April 2022, the first version of this online calculator is being used by 330 researchers and students with 26\,\% of them coming from the industry \cite{21}.

\begin{figure}[!ht] 
	\captionsetup{format=sanslabel}
    \hbox to\hsize{\hss\includegraphics[width=16cm]{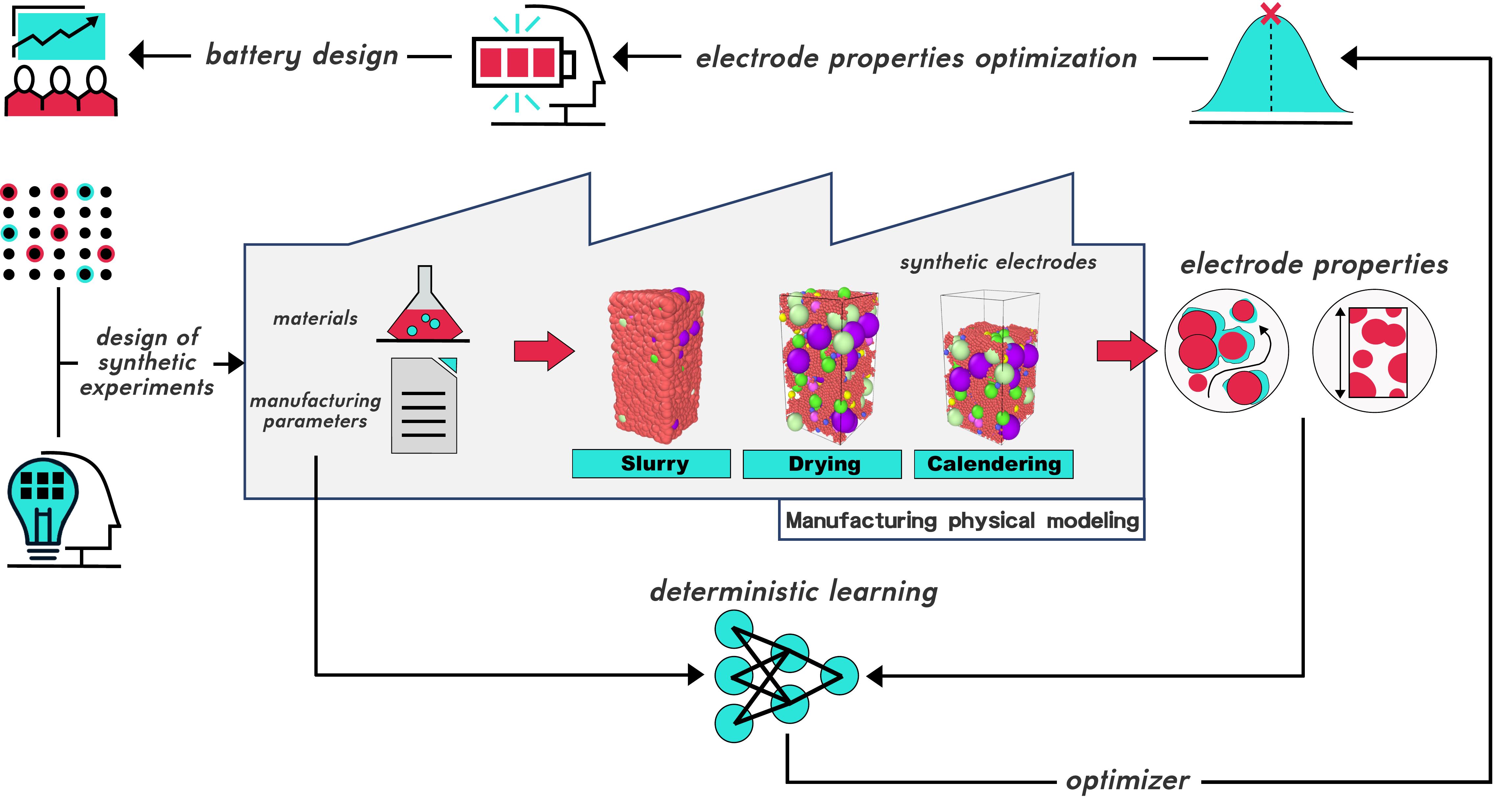}\hss} 
    \caption{ \textbf{Figure 1} : Schematic representation of our deterministic-assisted multi-objective optimization of electrode properties to design better lithium-ion batteries. The manufacturing process modeling from the slurry to the calendering is carried out through a chain of interlinked 3D physics-based models for the prediction of electrode mesostructures whose textural properties are evaluated. The coupling of a low discrepancy manufacturing parameters sequences with deterministic learnings allows bypassing the entire manufacturing physics-based modeling in the first step, then serving for the implementation of the optimization loop in the second step. The approach allows pinpointing the manufacturing parameters that have to be used in order to fabricate an electrode with optimal properties.}
\end{figure}

\paragraph{} While the ARTISTIC physics-based models are very useful for understanding and for predicting the influence of manufacturing parameters on the 3D electrode mesostructures and their electrochemical performance,\cite{22} they can have high computational cost (\textit{e.g.} several hours are needed for simulating an electrode slurry), hindering their usage to perform fast (in few seconds) electrode optimization. Such a fast optimization capabilities will be needed in digital twins collecting data through sensors, and giving instructions to the manufacturing machines through actuators for on the fly and autonomous optimization \cite{12,23,24,25}. In the ARTISTIC project, we have demonstrated that ML can also constitute a powerful tool to unravel correlations between manufacturing process parameters and electrode properties \cite{26,27,28,29}. ML has also demonstrated powerful capabilities to solve a wide diversity of scientific problems in the battery field \cite{30,31}. For instance, Tong \textit{et al.} applied Neural Networks for the prediction of LIB Remaining Useful Life,\cite{32} whereas Turetskyy \textit{et al.} used regression models to build a multi-output model predicting the final product properties of a battery production line.\cite{33} Furthermore, ML has been used to derive surrogate models of physics-based models describing LIB aging or redox flow battery operation \cite{34,35}. These models reproduce the prediction capabilities of the physics-based models with much cheaper computational costs. Furthermore, algorithms like the Bayesian Optimization (BO) framework which is supported on a probabilistic approach, constitutes powerful tools to solve optimization problems and perform inverse design \cite{36,37,38,39}. BO has seen widespread application in different domains such as cognitive science applications,\cite{40} autonomous driving,\cite{41} and pharmaceutical product development \cite{42}. Furthermore, Jiang \textit{et al.} quantified the decrease of a LIB charging time for single and multi-constant-current-step charging profiles,\cite{43} while Wang \textit{et al.} used BO for chemical products discovery and materials modeling \cite{44}. As the best of our knowledge, such an optimization workflow has never been used before to deal with the multi-objective optimization of battery electrodes and the prediction of the process parameters to adopt in order to manufacture the optimal electrodes.

\paragraph{} In this work, we report an innovative computational tool able to optimize multiple electrode properties at the same time, and able to evaluate the process parameters to adopt in order to manufacture them. We first generated a synthetic dataset containing inputs/outputs from a stack of ARTISTIC physics-based models simulating a NMC111-based electrode manufacturing process, using low discrepancy quasi-random sequences \cite{45}. This enabled us to define a batch of sparse manufacturing conditions, to then evaluate the properties of the various digitally  generated electrode mesostructures. The result constitutes a meaningful dataset for the training of deterministic learnings, whose main goal is to approximate each expensive electrode physics-based generation as a numerical function of the manufacturing parameters. These so-derived surrogate models cover the overall available manufacturing parameters space and bypass the manufacturing physical modeling chain as illustrated in Figure 1. Secondly, we took advantage of combining these models to raise a deterministic-assisted objective function for maximizing/minimizing selected properties simultaneously. This is carried out by embedding the latter function into a BO framework for this multi-objective optimization purpose. Our synthetic dataset accounts for the slurry, its drying, and the resulting electrode calendering, for the determination of the best amount of active material (AM\%), slurry solid content (SC\%), and the calendering compression rate (CR\%), giving an electrode with minimal tortuosity factor, maximal effective electronic conductivity, maximal active surface area between AM and pores, and maximal density. Such optimized properties are expected to influence the kinetic, ionic, and electronic transport properties of the electrodes. The so-found best manufacturing condition was finally used to manufacture a real electrode to experimentally validate the relevance of our workflow prediction. In the following, we explain our approach in detail, we present the results and we explain why we believe this work paves the way towards the emergence of autonomous battery manufacturing optimization procedures.

\subsection{Low Discrepancy Sequences for the Design of Experiments}
\paragraph{} Given the high computational cost of the manufacturing process physics-based modeling, it is time and resources consuming to generate data which are highly-representative of the entire manufacturing parameters space. It is therefore important to find a computationally cheaper and automatic way to probe such a space (called hereafter the inputs space). To do so, we generate a design of experiments (DOE) based on highly-representative manufacturing parameters values, which were used as inputs of the physics-based models for the data generation (electrode mesostructures and associated properties) \cite{46}. This was done through the implementation of low discrepancy quasi-random Sobol sequences with Saltelli extension (Figure 2) \cite{47}. Such sequences sample quasi-uniform distributions of data from continuous input parameters values, and probe in a proper manner to accurately capture the inputs space, compared to usual uniform or Gaussian distributions \cite{48}. Figure 2A displays a schematic representation of a 2D-generation of data based on the previously mentioned distributions, supporting the choice of Sobol sequences to design highly-representative synthetic manufacturing parameters values. In our case study, we used as input parameters the AM\% and the SC\% related to the slurry step, and the CR\% related to the calendering step, and then selected fixed boundaries for each of them, to form the inputs space illustrated by the hyperrectangle in Figure 2B. \\

\begin{figure}[!ht] 
	\captionsetup{format=sanslabel}
    \hbox to\hsize{\hss\includegraphics[width=14cm]{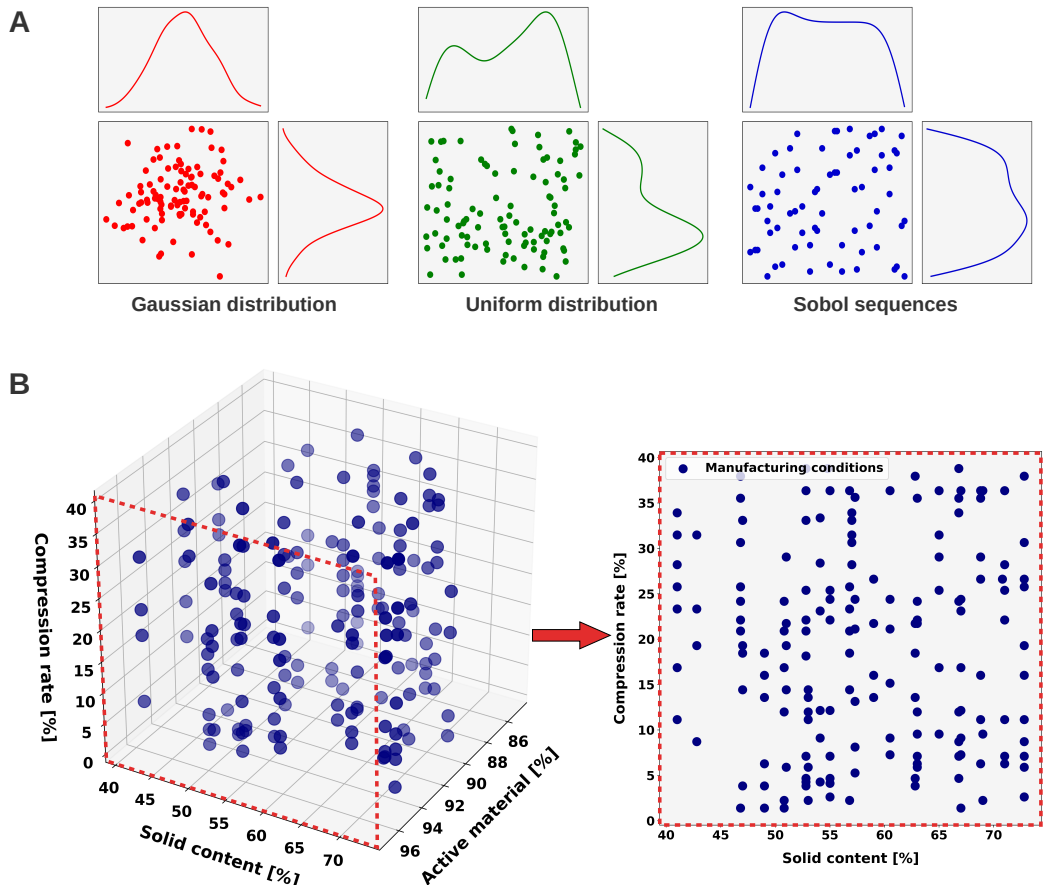}\hss} 
    \caption{ \textbf{Figure 2} : \textit{(A)} From the left to the right, schematic representation of 2D data generation with Gaussian distributions, uniform distributions, and Sobol sequences. The empirical distribution was added for each dimension to display the representativity of values within the 2D space; \textit{(B)} Design of experiments for the synthetic dataset through the Sobol sequences after the sampling modifications within our inputs space. The 3D representation is straightforward due to a space-filling hyperrectangle design. A 2D representation of the generated sequences based on the CR\% values as a function of SC\% values displays the quasi-randomness of the input space-filling.}
\end{figure}

\paragraph{} In our manufacturing physics-based modeling workflow, the generation of a slurry and its corresponding dried electrode is computationally expensive (\textit{e.g.} $\sim$ 150 hours), while the generation of a calendered electrode is drastically faster (\textit{e.g.} $\sim$ 8 hours). As a consequence, we proceeded to a post-treatment of the quasi-random sequences regarding the variables AM\%, and SC\% (inputs of the slurry physics-based modeling) to tackle the overall high computational cost. This post-treatment consisted in selecting a limited amount of slurry manufacturing parameters values (AM\%; SC\%) among those generated in the DOE and associating them randomly to the evaluations of CR\% values generated by the quasi-random sequences. In the end, our DOE corresponded to a shortened quasi-random generation of diverse slurries (and associated drying) where different quasi-random calendering conditions can be associated with one specific electrode slurry. In particular, this DOE was shortened to reduce the computational time when generating the associated 3D mesostructures whereas its representativity from the inputs space was meaningful as displayed in Figure 2. According to Cervellera \textit{et al.}, this strategy of sequences generation widely influences the results in ML applications and optimization processes \cite{49}.

\subsection{Data Acquisition}
\paragraph{} For each set of manufacturing parameters values (AM\%, SC\%, CR\%) in the DOE, we generated the different 3D NMC111 electrode mesostructures by using the ARTISTIC physics-based models simulating the slurry, the drying, and the calendering of the resulting electrodes \cite{13,17,18}. These models were developed under LAMMPS software \cite{50}. We processed the mesoscale simulated Nickel-Manganese-Cobalt (NMC111)-based slurry, supported by the Coarse-Grained Molecular Dynamics (CGMD) \cite{13}. This simulation encompasses Force-Field parameters (FFs) to describe the physicochemical interaction between active material (AM) and carbon-binder domain (CBD) phases during the slurry equilibration \cite{13}. Then, the dried electrode mesostructure was obtained by uniformly shrinking the CBD particles to mimic the solvent removal and get the equilibrated electrode mesostructure \cite{51}. With the obtained dried mesostructure, we proceeded to the calendering of the resulting dried electrode by using the DEM which simulates its mechanical behavior \cite{52,22}. The resulting calendered electrode mesostructures were characterized in terms of thickness, volumes, densities, and active surface areas between AM and pores using in-house Python scripts, their effective electronic conductivities using GeoDict,\cite{53} and their tortuosity factors of the pore network using TauFactor \cite{54}. For all the electrode generation, we fixed the total mass of the electrode equal to 0.1 $\mu$g.

\subsection{Deterministic Learnings}
\paragraph{} The calculated 3D electrode properties versus manufacturing parameters values constitute a meaningful database for training deterministic learnings (\textit{i.e.} regression function trained on a pre-defined dataset). Indeed, we replaced the entire manufacturing physics-based modeling chain with a surrogate model to directly calculate the electrode properties as a function of the manufacturing parameters. Since the synthetic dataset covers well the inputs space, the latter learnings can accurately calculate the electrode properties regardless of the manufacturing conditions \cite{55}. This was concluded thanks to the training and testing of the \textit{Sure Independence Screening and Sparsifying Operator} (SISSO) algorithm, according to good capabilities to interpolate within the inputs space. SISSO allows to directly obtain a mathematical equation interlinking electrode properties with manufacturing parameters, something convenient for proceeding into the optimization process \cite{56,57,58}. The algorithm mentioned above mainly provides a linear relationship between an output y and a set of descriptors $(d_{i})_{i \leq n}$ formed by non-linear relationships between inputs, defined as

\begin{ceqn}
\begin{align}
\tag{Eq. 1}
y = \sum_{i=1}^{n}  c_i \times d_i \quad c_i \neq 0 
\end{align}
\end{ceqn}

\paragraph{} The equation above lies on two pillars: a first one based on the feature space construction for descriptors implementation,\cite{59} and a second one on a solution algorithm for the error minimization \cite{60,61}. The validation of each deterministic learning per electrode property was achieved through relevant validation metrics discussed in the following section.

\subsection{Optimization Process}
\paragraph{} We proceeded to a Bayesian multi-objective optimization to assess an objective function (denoted \textit{$C_f$}), itself dependent on different deterministic learnings. We built a scalarizing function after transforming the multi-objective problem into a single objective function. In that sense, one of the popular ways of applying BO for multi-objective optimization problems is using Gaussian Process (GP) regressions as the model to approximate \textit{$C_f$}. In each iteration, BO calculates a posterior distribution \textit{$C_f$} $| D$ over \textit{$C_f$} regarding the set of all previous data $D$. Then, this proposes a new set of manufacturing parameters values, chosen by an acquisition function, balancing between the exploitation of prior parameters values combinations to identify nearby minima, and the exploration to identify minima far from prior parameters values combinations \cite{62,63}. The GP model is updated at each step until the BO returns the manufacturing parameters values giving origin to an electrode with the optimal properties \cite{37,64}. In this study, we assessed a minimization problem to figure out the optimization of the electrode electrochemical performance and electronic/ionic transport efficiency, by formalizing the search for the best manufacturing parameters values  giving the best electrode as it follows:

\begin{ceqn}
\begin{align}
\tag{Eq. 2}
    x^{*} = argmin_x(C_f(x))
\end{align}
\end{ceqn}

\paragraph{} We maximized simultaneously the effective electronic conductivity, the density, and the active surface area between AM and pores, the latter being assumed to be fully filled with the electrolyte, and we minimized the tortuosity factor. This multi-objective optimization was done by granting equal weights to each property. Due to different orders of magnitudes, we adopted a scalarization method for \textit{$C_f$} through the incorporation of a scalar fitness function, allowing to fit each property value into [0, 1] when calculating the objective function values. This scalarization avoids a bias induced by the properties values, balancing the objective function between the maximization of some of the properties and the minimization of the others \cite{37}. Each scaled electrode property (denoted $y_{i,x}$ $\forall i \leq 4$) was included in \textit{$C_f$} in order to minimize $y_{i,x}$ when the associated property used to be minimized, and to minimize $1-y_{i,x}$ when the property used to be maximized. In this regard, we proposed a \textit{$C_f$} satisfying the constraints by an equal-weighted power function as it follows \cite{64}

\begin{ceqn}
\begin{align}
\tag{Eq. 3}
    C_f = \frac{1}{4}(\sum_{y_{i,x} \subset Y_m} y^2_{i,x} + \sum_{y_{i,x} \subset Y_M} (1-y_{i,x})^2)
\end{align}
\end{ceqn}

\noindent where $Y_m$ and $Y_M$ define respectively the set of electrode properties that are minimized and maximized in our study.

\paragraph{} In addition, \textit{$C_f$} was built with mathematical equations from each deterministic learning, predicting the property as a function of a specific set of manufacturing parameters values x. Consequently, the objective function calculations were a quick non-linear combination of predictions without the need to run the entire manufacturing process physics-based modeling chain. This presents the advantage of a reduced computational cost at assessing the multi-objective optimization process, and then obtaining the best manufacturing parameters values associated with the optimal electrode.

\section{Results}
\subsection{Validation of the learnings}
\paragraph{} The root mean square error in percentage (RMSE\%) and the $R^2_{score}$ for the goodness of fit, were chosen as validation metrics to evaluate and validate the training/testing of the different deterministic learnings. As a standard practice in ML, the whole dataset has been divided randomly into training and testing datasets for model training and validation, respectively \cite{65}. The training dataset contained 80\,\% of data, randomly picked up from the synthetic dataset which contains 174 sets of manufacturing parameters values in total, and the testing dataset contains the remaining 20\,\%. In Table 1, we report the values for the testing dataset, where the results are supported by the regression plots displayed in Figure S1 in the Supplementary Information. Moreover, we also report the values for the training dataset in Table S1  in the Supplementary Information. Even though we obtained very high $R^2_{score}$ values and low RMSE\% values, we determined the 95\,\% confidence interval (CI95) for one of metrics: the $R^2_{score}$ \cite{66}. The latter interval provides more statistical analytics on the capabilities of the SISSO algorithm to train/test regardless of the seed of the re-sampling procedure. Indeed, confidence intervals estimate the variability of the metrics on how precise it is likely to be. Therefore, there is a 95\,\% likelihood that our CI95 covers the ranges detailed in Table 1, reporting the true learning performance \cite{67}.

\begin{table}[h]
\centering
\small
\setlength\tabcolsep{3pt}
\setlength\extrarowheight{2pt}
\captionsetup{format=sanslabel}

\begin{tabularx}{\textwidth}{ 
  >{\raggedright\arraybackslash}X 
  >{\raggedright\arraybackslash}X
  >{\raggedright\arraybackslash}X
  >{\raggedright\arraybackslash}X}
\midrule
\textsc{Property}   & \textsc{RMSE\%}   & \textsc{$R^2_{score}$}  & \textsc{CI95} \\ \midrule
Electronic conductivity [S/m]      &  1.48 & 0.933 & [0.941; 0.950] \\
Tortuosity factor      & 7.80 & 0.979 & [0.978; 0.981]\\
Active surface area [\%]      &  1.41 & 0.911 & [0.885; 0.914]\\
Density [g/cm$^3$]      &  1.87 & 0.968 & [0.962; 0.971]\\
    \midrule          
\end{tabularx}
\caption{\textbf{Table 1} : Validation metrics calculated over the testing dataset, and associated to the fitting of the electrode properties. The 95\,\% confidence interval (CI95) were estimated with a total of 75 random seeds of training/testing datasets for the uncertainty of the $R^2_{score}$.}
\end{table}

\paragraph{} The results showed high predictive capabilities of the different deterministic learnings to predict the electrode properties only by giving the three manufacturing parameters as inputs. The CI95s did not have extended boundaries for the selected 95\,\% likelihood, with an RMSE\% very low, suggesting that the synthetic dataset did not provide a bias when setting the seed and testing the deterministic learnings. This suggests that the trained deterministic learnings were suitable regression models to predict the kinetic, ionic, and electronic transport properties, thus bypassing the entire manufacturing physics-based modeling chain.

\subsection{Optimal manufacturing parameters values}
\paragraph{} The BO framework was designed with 300 iterations as a cut-off to propose a meaningful candidate for the minimization of the objective function. Figure 3 shows the 2D partial dependence plot, allowing to interpret the importance of the different manufacturing parameters in the GP model. The principle lies in the visualization of the marginal effect of manufacturing parameters on the approximation of \textit{$C_f$} values, given an average influence of all the other parameters. Since the input parameters are uncorrelated, the 2D partial dependence plot constitutes a tool to interpret how predictions change when fixing one manufacturing parameter value every time and by freeing the others two \cite{68}. \\

\begin{figure}[!ht] 
	\captionsetup{format=sanslabel}
    \hbox to\hsize{\hss\includegraphics[width=18cm]{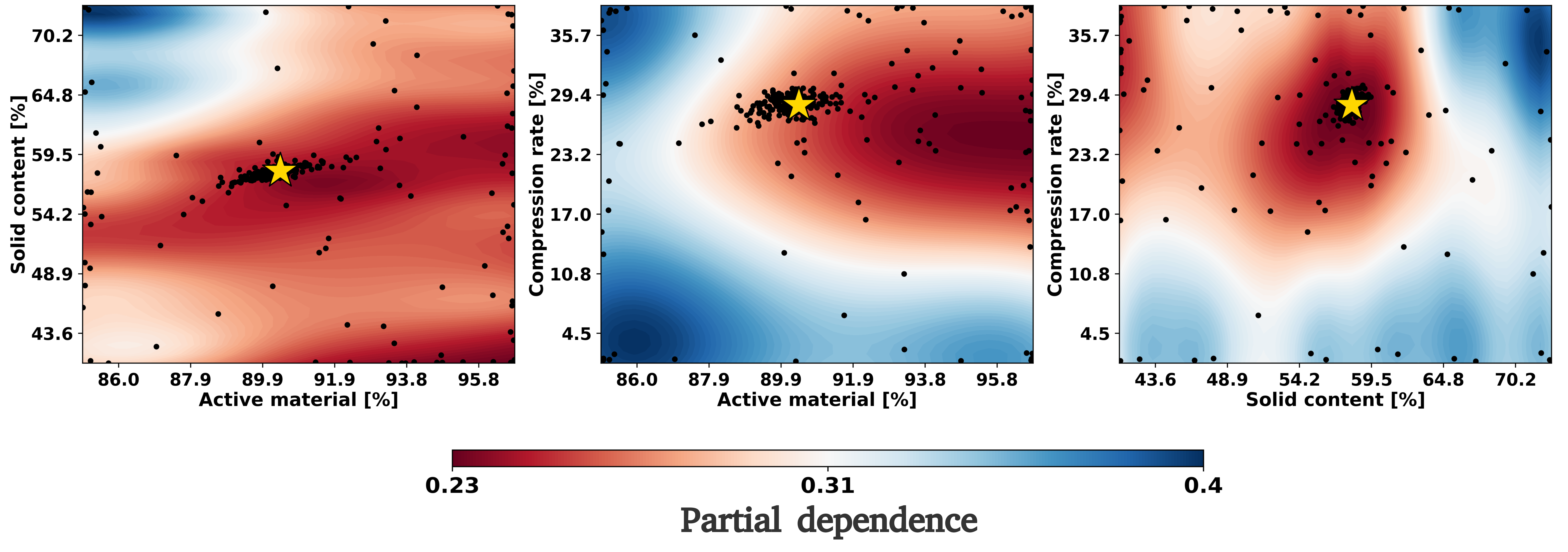}\hss} 
    \caption{ \textbf{Figure 3} : Partial dependence plots allowing to interpret the GP model’s predictions. A 2D representation was straightforward to better visualize how manufacturing parameter values influence Cf values for the search of the optimal condition. The results are color-coded where warmed values suggest less variability in the predictions when changing the hidden input parameter values, contrary to cooler values which suggest higher variability. The yellow stars point out the pairwise optimal manufacturing conditions predicted by the BO framework.}
\end{figure}

\paragraph{} Table 2 reports the best manufacturing parameters predicted by the BO framework which balancing the properties values for the initial multi-objective optimization problem.

\begin{table}[h]
\centering
\small
\setlength\tabcolsep{3pt}
\setlength\extrarowheight{2pt}
\captionsetup{format=sanslabel}

\begin{tabularx}{\textwidth}{ 
  >{\raggedright\arraybackslash}X 
  >{\raggedright\arraybackslash}X
  >{\raggedright\arraybackslash}X}
\midrule
\textsc{Active material [\%]}   & \textsc{Solid content [\%]}   & \textsc{Compression rate [\%]} \\ \midrule
90.4 & 58.2 & 28.4 \\
    \midrule          
\end{tabularx}
\caption{\textbf{Table 2} : Optimal manufacturing parameters predicted by the BO framework.}
\end{table}

\paragraph{} As a result, it can be seen that the optimal solution (yellow star) was found in the region of SC\,\% and CR\,\% where the partial dependence is the lowest. In contrast to this result, the influence of AM\,\% values is more independent at higher values of AM\,\% (> 92\,\%), suggesting a less straightforward marginal effect of AM\,\%. Nevertheless, Figure 3 displays that the aggregation of sets of manufacturing parameters values tested by the BO framework, falls into the region of the inputs space where the optimal solution is located. This highlights the rapidity of the BO to find the region of the best candidate, as it can be also seen through the convergence plot from Figure S2 in the Supplementary Information. In such a way, we were able to obtain the best values of the three manufacturing parameters that optimize the electrode properties of interest simultaneously and that do not represent the extreme values of the manufacturing parameters ranges used in the BO framework.

\section{Discussion}
\subsection{Optimized transport properties}

\begin{figure}[!ht] 
	\captionsetup{format=sanslabel}
    \hbox to\hsize{\hss\includegraphics[width=16cm]{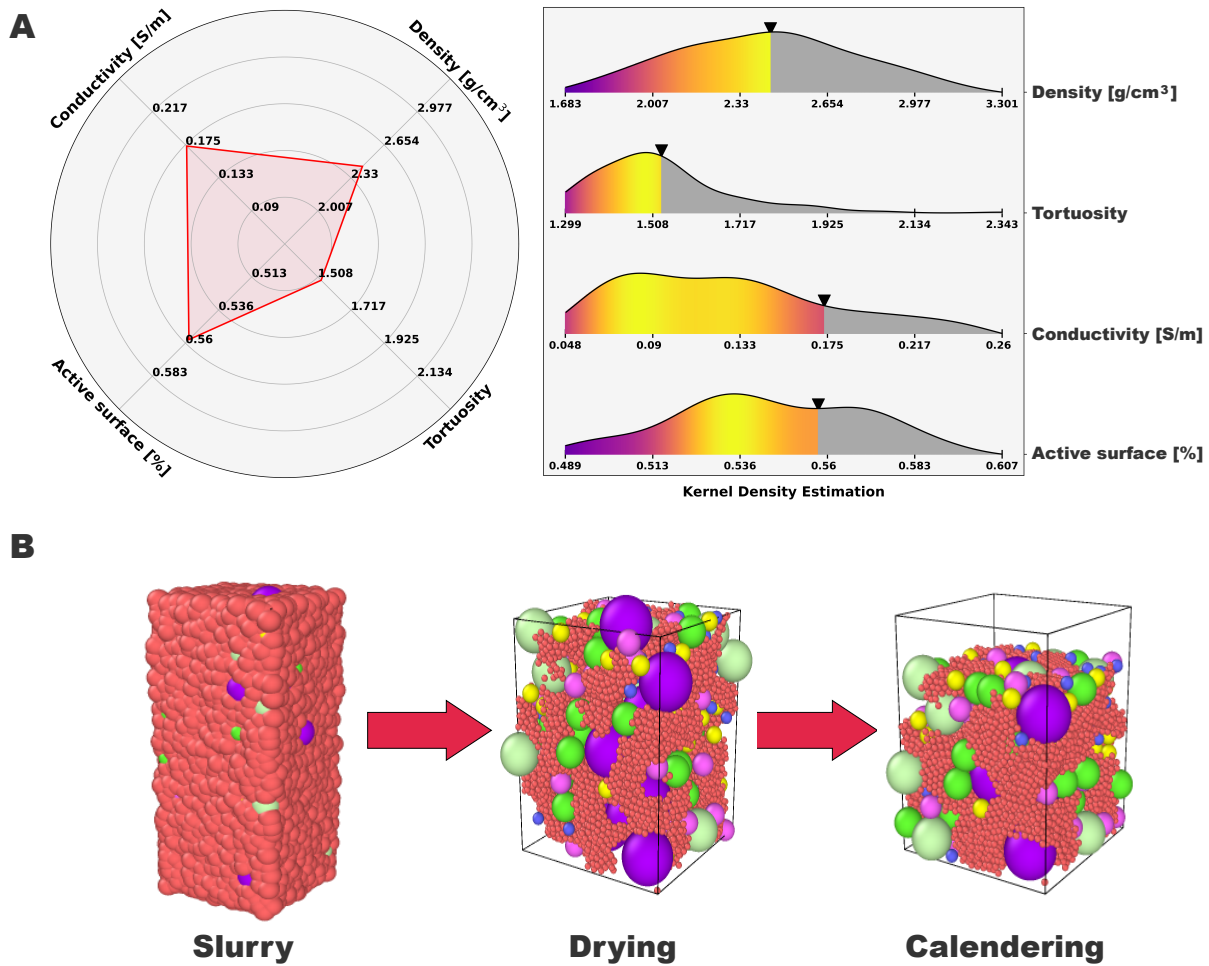}\hss} 
    \caption{ \textbf{Figure 4} : \textit{A)} Graphical representation of the optimized electrode properties obtained through the deterministic-assisted optimization loop. A radar chart displays the optimized values after generating the 3D electrode mesostructures using the optimal manufacturing condition. The values have been scaled between the min and max for each property from the synthetic dataset. A Kernel Density Estimation was added in order to display how the optimized values are represented within the empirical distribution of properties from the synthetic dataset. A gradient color-codes such a distribution until the optimized value. We replaced electronic conductivity by conductivity, tortuosity factor by tortuosity, and active surface area by active surface in this Figure; \textit{(B)} 3D electrode mesostructures from the slurry to the calendering step associated to the optimal manufacturing condition (Table 2) predicted by the deterministic-assisted optimization loop.}
\end{figure}

\newpage
\paragraph{} The four output properties (electronic conductivity, active surface area, tortuosity factor and density) correspond to different electrochemical characteristics during the LIB electrode operation. For instance, the electronic conductivity affects the solid phase electrostatic potential gradient. The active surface area affects the reaction kinetics and the associated current density on the surface of the AM. Under a certain terminal current, a higher active surface area means a lower reaction current density. According to the Butler-Volmer equation, the overpotential of reaction is reduced \cite{69}. The tortuosity factor is related to the effective diffusivity of Li+ in the electrolyte and ionic conductivity. The lower the tortuosity factor is, the easier it is for ions to transport. The density is vital for the evaluation of the areal energy density of the electrode. To achieve the highest energy density, the electrode should have the highest active surface area, the highest electronic conductivity and the highest density, with a tortuosity factor equal to 1. However, in practice, these four properties are related to each other. Therefore, an optimized electrode will result from a balance between the values of these properties, with the best performance achievable under the constraint of the manufacturing parameter space.

\paragraph{} Figure 4 displays the four properties values of the optimal electrode obtained by our BO framework. The Figure 4 also displays the corresponding 3D electrode mesostructure, calculated using the manufacturing physics-based modeling chain by taking as inputs the manufacturing parameters values of the optimal electrode, and which has a porosity equal to 29.4\,\%. The Kernel Density Estimation (KDE) for each property reflects the localization of the optimal electrode properties values compared to the empirical distribution of values from the synthetic dataset. This allows emphasizing the representativity of a high-performance case compared to the values of the different distributions. For instance, we noticed an optimized conductivity of 0.175 S/m, while the corresponding KDE reflected low representativity of higher values within the synthetic dataset.

\paragraph{} In Figure 5, we compare the optimal case with four extreme cases of electrodes from the synthetic dataset, each of them having the highest performance for only one property. In the first case (blue vs. red), the electronic conductivity is pushed to the extreme by increasing the content of the CBD phase and then reducing AM\,\%. This leads to a low electrostatic potential gradient in the solid (AM+CBD). In the meantime, more AM surface area is covered by the CBD phase. According to previous researches, the CBD phase is a porous conductive matrix, with a porosity of approximately 50\,\% \cite{70}. The low degree of exposure of AM to the pores will increase the surface reaction current density, triggering a higher Li$^{+}$ (de-)intercalation reaction overpotential. Furthermore, the effective electronic conductivity and diffusivity within the CBD phase were found to be 5\,\% of the bulk phase \cite{71}. Therefore, this result in a lower effective diffusivity in the whole electrode. The optimal case, on the contrary, has a higher active surface area and density, which is in favor of obtaining a higher energy density.

\paragraph{} Case 2 (green vs. red) illustrates the comparison with the tortuosity factor-optimized electrode. The low tortuosity factor results from the low compression rate of this electrode (2.24\,\%). From a manufacturing process viewpoint, this case is close to an uncalendared electrode. It is worth to be noticed that the electronic conductivity drops due to the poor connection between the CBD particles. Electrode capacity can be severely undermined without a well-established electronic conductive network. While increasing electrostatic potential drop in the solid phase can also cause particles isolation and inactivation. Case 3 (yellow vs. red) aimed at the high active surface area. From the manufacturing process perspective, this is the case where solid content is rather high (60.3\,\%), which is a favorable way to increase electrode energy density. The cost of raising density while maintaining a porosity of 40\,\% is having reduced CBD content. Compared to the optimal case, the conductivity is more than two times lower. Case 4 (magenta vs. red) exhibits the highest density due to a large compression rate (36.36\,\%). The highly compact structure dramatically increases the tortuosity factor and decreases the active surface area. In reality, there is a risk of breaking AM particles alongside the increased overpotential, something which is not considered in the current version of our optimization workflow. The newly generated surface of the crack in the AM particles can cause extra parasitic reactions, such as Cathode Electrolyte Interface CEI growth and transition metal dissolution. The pulverized particles can also lose connection with the conductive network, therefore becoming inactive. Our multi-objective optimization pipeline allows to suggest an optimal electrode for moderate calendering conditions typically used in battery prototyping lines.

\paragraph{} It is worth to notice that our optimal electrode result is acquired based on the weight we put on each property according to Eq. 3. Depending on different electrode systems or manufacturing conditions, the weight can be further adjusted, resulting in different optimal cases regarding the final battery cell application from an industrial perspective (\textit{e.g.} fast-charging and high power applications, cell durability, ...). Indeed, the real battery cell performance depends on how we adjust the electrode properties to assess the best electrode properties. In that sense, our optimization workflow can be generalized to any kind of optimization problem for different applications. 

\newpage 
\begin{figure}[!ht] 
	\captionsetup{format=sanslabel}
    \hbox to\hsize{\hss\includegraphics[width=16cm]{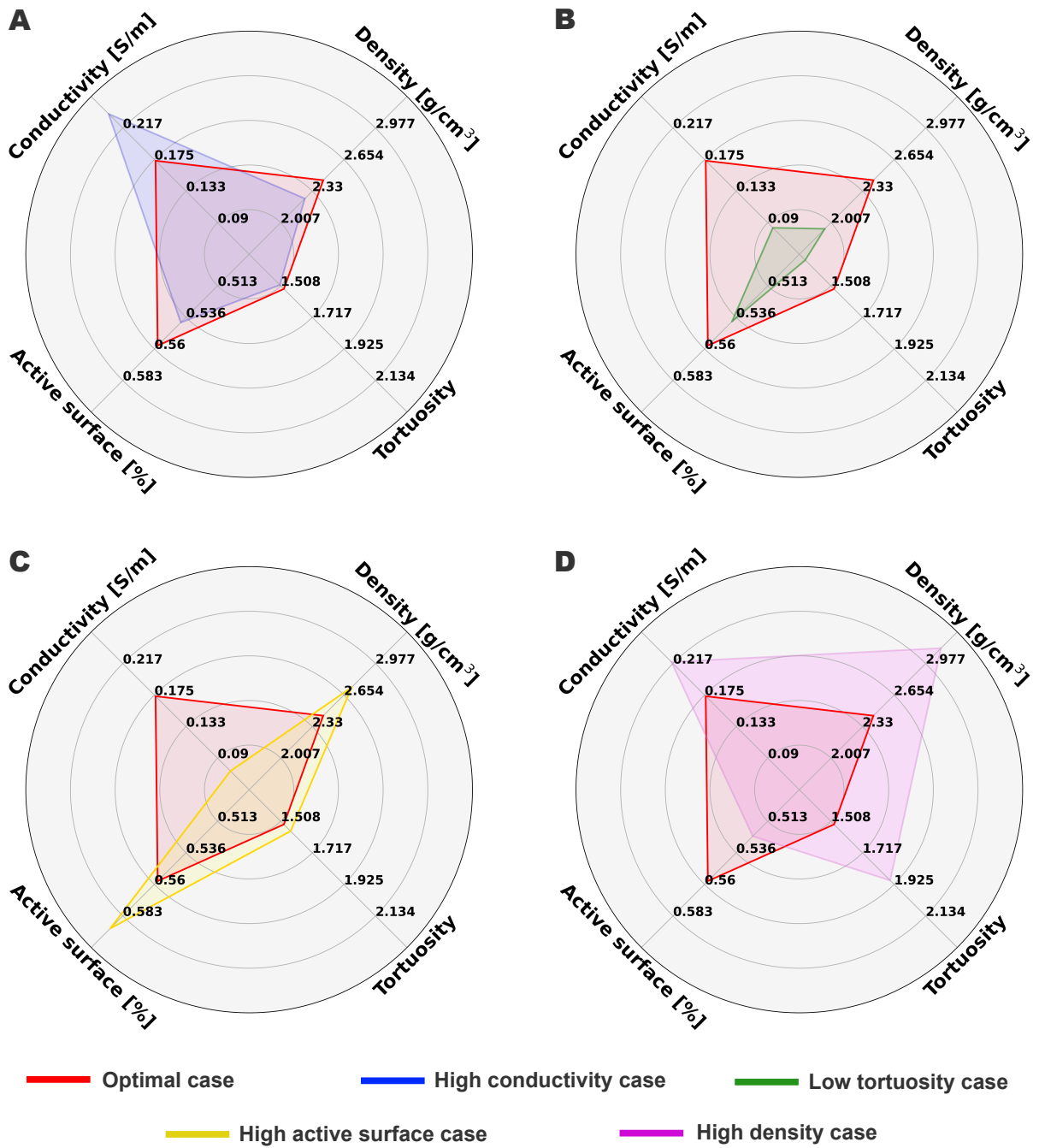}\hss} 
    \caption{ \textbf{Figure 5} : Radar chart plots comparing the optimized electrode properties with extreme cases from the synthetic dataset. Each plot shows the interest to obtain the optimal case (red) instead of having a high-performance electrode for only one property. We replaced electronic conductivity by conductivity, tortuosity factor by tortuosity, and active surface area by active surface in this Figure; (A) High electronic conductivity case ((AM\%, SC\%, CR\%) = (86.3\,\%, 57\,\%, 33.08\,\%)) with a porosity of 26.55\,\%; (B) Low tortuosity factor case ((AM\%, SC\%, CR\%) = (88.8\,\%, 56.8\,\%, 2.24\,\%)) with a porosity of 42.35\,\%; (C) High active surface area case ((AM\%, SC\%, CR\%) = (96.5\,\%, 60.3\,\%, 9.11\,\%)) with a porosity of 35.97\,\%; (D) High density case ((AM\%, SC\%, CR\%) = (94.5\,\%, 71\,\%, 36.36\,\%)) with a porosity of 24.11\,\%.}
\end{figure}

\newpage 
\subsection{Experimental relevance of the predicted optimal electrode} 
\paragraph{} In order to assess the experimental relevance of the optimal electrode predicted by our multi-objective optimization pipeline (Figure 4), we used the corresponding manufacturing parameters values to manufacture such an electrode by using our battery prototyping line. To avoid calibration errors in the prototyping machines, we have rounded the manufacturing parameters values suggested by our BO framework. Even though our 3D physics-based models did not take into account all the manufacturing parameters that can be tuned in the prototyping machines (\textit{e.g.} slurry drying temperature), we have adjusted the coating comma gap and the coating speed, and the temperatures of the two-part oven of our prototyping line in order to fix meaningful values to match the optimized electrode properties proposed by our BO framework (see the Methods section). The resulting experimental properties are reported in Table 3, and showed a reasonable agreement while compared to the modeling results reported in Figure 4. This highlights the relevance of our approach to give a practical application from the manufacturing physics-based modeling optimization, suggesting new designs of high-performance cells.

\begin{table}[h]
\centering
\small
\setlength\tabcolsep{3pt}
\setlength\extrarowheight{2pt}
\captionsetup{format=sanslabel}

\begin{tabularx}{\textwidth}{ 
  >{\arraybackslash}X 
  >{\arraybackslash}X
  >{\arraybackslash}X}
\midrule
\textsc{Property}   & \textsc{Pristine}   & \textsc{Calendered 30\,\%]} \\ \midrule
Mass loading [mg/cm$^2$] & 6.7 $+/-$ 0.3 & 6.7 $+/-$ 0.3 \\
Thickness [$mu$m] & 68 $+/-$ 4 & 48 $+/-$ 4 \\
Density [g/cm$^3$] & 1.6 $+/-$  0.4 & 2.6 $+/-$ 0.3 \\
Tortuosity factor & 3.5 $+/-$ 0.5 & 1.8 $+/-$ 0.4 \\
Porosity [\%] & 0.60 $+/-$ 0.03 & 0.29 $+/-$ 0.07 \\
    \midrule          
\end{tabularx}
\caption{\textbf{Table 3} : Properties of the cathode electrode experimentally manufactured with the manufacturing parameters values evaluated as optimal by the optimization pipeline. Tortusoity factors were calculated by using the Transmission Line method proposed by Landesfeind \textit{et al} \cite{72}.}
\end{table}

\section{Conclusions}
\paragraph{} In this study, we reported an innovative deterministic-assisted multi-objective optimization approach of different LIB electrode properties (related to kinetics, electronic and ionic transport) simultaneously. Such an approach predicts the process parameters to adopt in order to manufacture the so-found optimal electrode. To do so, we took advantages of the previously developed 3D-resolved ARTISTIC physics-based electrode manufacturing models simulating the slurry, its drying, and the resulting electrode calendering, to raise a synthetic dataset that contains representative parameters of these manufacturing steps with various resulting electrode properties. The application of low-discrepancy sequences enabled us to generate a shortened but sufficient batch of manufacturing parameters values to successfully implement a deterministic learning-derived surrogate model embedded within a multi-objective optimization loop. The surrogate model allows capturing the impact of manufacturing parameters on the electrode properties with the advantage of being computationally cheaper (few seconds for the predictions) than the physics-based modeling chain. This characteristic enables a faster minimization of the objective function by using a BO framework. Furthermore, we showed that the tortuosity factor, the electronic conductivity, the active surface area, and the density are bound to each other simultaneously, with the highest electrode performance is expected under balanced properties. 

\paragraph{} Last but not least, we have manufactured the electrode predicted by our framework as the optimal one, in order to experimentally validate this prediction relevance. As a perspective, we aim to extend our study to other manufacturing steps, such as the electrolyte infiltration, the formation and the electrochemical performance. The overall proposed approach in this article, while being demonstrated for LIBs, can be transferred to the manufacturing of other battery technologies and the manufacturing of composite materials in general \cite{73}. In addition, we believe that our approach can be adapted to optimize LIB performance and lifetime, through the generation of synthetic data from physics-based performance models accounting for multiple aging mechanisms (\textit{e.g.} SEI formation, lithium plating). This can lead to new ways of analyzing LIB degradation mechanisms to increase its performances in terms of energy, power density and durability. Finally, we believe that our approach also paves the way towards the design of hardware/software infrastructures allowing to perform autonomous optimization in battery manufacturing processes. Such infrastructures could be built on top of experimental data on-the-fly acquisition (\textit{e.g.} electrode thickness), data augmentation through the use of the ARTISTIC physics-based models to generate highly-representative synthetic data, the deterministic learnings to derive surrogate models to evaluate the influence of process parameters to adopt being sent as instructions to actuators tuning on the fly the machines within the production or prototyping line manufacturing the electrodes. Such a futuristic vision has the potential to transform the way we optimize battery manufacturing processes, accelerating the energy transition of our societies.

\section*{Methods}
\subsection*{Electrode processing}
\paragraph{} In order to prepare the electrode, we used LiNi$_{1/3}$Mn$_{1/3}$Co$_{1/3}$O$_2$ (NMC111), active material supplied by Umicore. We employed C-NERGYTM super C45 carbon black (CB) supplied by IMERYS. SolefTM Polyvinylidene fluoride (PVDF) was used as a binder and purchased from Solvay, and N-methyl pyrrolidone (NMP) was used as a solvent from BASF. The slurry components (90\,\% NMC111, 6\,\% CB, and 4\,\% PVDF) were premixed with a soft blender. Afterward, NMP was added until reaching a desired solid content (SC) of 57\,\%, a ratio between the solid components and the solvent. The mixture was performed in a Dispermat CV3-PLUS high-shear mixer for 2 hours in a water-bath cooled recipient at 25 $ºC$. The slurry was coated over a 22 $\mu$m thick Aluminum current collector using a comma-coater prototype-grade machine (PDL250, People \& Technology, Korea), fixing the gap at 90 $\mu$m and the coating speed at 0.3 $m/min$. The electrodes were dried in a built-in two-part oven at 80 and 95 $ºC$. The electrodes were calendered with a prototype-grade lap press calender (BPN250, People \& Technology, Korea). The latter consists of a two-roll compactor of 25 cm in diameter. The gap between the rolls was set at 37 $\mu$m to reach 30\,\% of compression. The calendering was performed at constant line speed (0.54 $m/min$) and 60 $ºC$. The properties of the electrodes are presented in Table 3. EIS tests for calculating the tortuosity were performed in 2035 coin cells assembled in a dry room (H2O < 15 $ppm$). The coin cells (both on the positive and negative side) were assembled using Celgard 2500 as separator (thickness = 25 $\mu$m, porosity = 55\,\%, mass = 2.25 $mg$), NMC cathodes (diameter = 13 $mm$, mass = 16.4 $+/-$ 0.2 $mg$ ), positive and negative casing (mass = 0.8715 and 0.8606 g, respectively), two current collectors (thickness = 0.5 and 1.0 $mm$, mass = 0.758 and 1.541 g, respectively), and a spring (mass = 0.1780 g). The electrolyte was a 10 mM TBAClO4 solution, prepared in a 1:1 wt mixture of ethylene carbonate:dimethyl carbonate (volume = 100 $\mu$L, mass = 0.148 g). The total battery mass was 4.33 $+/-$ 0.01 g. The EIS tests were performed with an MTZ-35 impedance analyzer (BioLogic, Seyssinet-Pariset, France) in 10$^{-1}$ - 10$^{7}$ $Hz$ with a potential perturbation of 5 mV. All measurements were carried out at 25 $+/-1$ 1$ºC$. The effective electronic conductivity was not characterized due to inapt experimental conditions. The measurements of ionic conductivity require specific experimental procedures for obtaining results in the liquid phase, which we were unable to attain, for instance, unsticking the current collector from the electrode and finding a suitable liquid phase. 

\subsection*{Validation Metrics}
\paragraph{} The root mean square error in percentage (RMSE\%) is defined by \\

\begin{ceqn}
\begin{align}
\tag{Eq. 4}
    RMSE\% = \sqrt{\frac{1}{n}\sum_{i=1}^{n} \frac{(y_i-\Tilde{y_i})^2}{\Tilde{y_i}^2}}
\end{align}
\end{ceqn}

\paragraph{} The $R^2_{score}$ is defined by \\

\begin{ceqn}
\begin{align}
\tag{Eq. 5}
    R^2_{score} = 1 - \frac{\sum_{i=1}^n (y_i-\Tilde{y_i})^2}{\sum_{i=1}^n (y_i-\bar{y_i})^2}
\end{align}
\end{ceqn}

where $y_i$, $\Tilde{y_i}$, and $\bar{y_i}$ are the predicted values by the deterministic learning, the real value from the dataset, and the average of the real values respectively.

\subsection*{Bayesian Optimization}
\paragraph{} Bayesian Optimization (BO) aims to minimize \textit{$C_f$} by approximating it with a Gaussian Process (GP) regression model, that takes into account a batch of input manufacturing conditions D = {x$_1$, x$_2$, x$_3$, ..., x$_k$} and the associated objective values \textit{$C_f$}(x$_1$), \textit{$C_f$}(x$_2$), \textit{$C_f$}(x$_3$), …, \textit{$C_f$}(x$_k$). Given the prior knowledge over  = [\textit{$C_f$}(x$_1$), \textit{$C_f$}(x$_2$), \textit{$C_f$}(x$_3$), ..., \textit{$C_f$}(x$_k$)], the BO assigns a multivariate Gaussian distribution \\

\begin{ceqn}
\begin{align}
\tag{Eq. 6}
    \bar{C_f} = GP(\mu_0(x_{1:k}), \Sigma_0(x_{1:k}, x_{1:k}))
\end{align}
\end{ceqn}

where $\mu_0(x_{1:k})$ and $ \Sigma_0(x_{1:k}, x_{1:k})$ are the expectation vector and covariance matrix respectively.

\paragraph{} Then, the GP model has to infer the posterior distribution \textit{$C_f$} | D which is assumed to follow a Gaussian distribution as \\

\begin{ceqn}
\begin{align}
\tag{Eq. 7}
    C_f | D = GP(\mu_*(x), \Sigma_*(x))
\end{align}
\end{ceqn}

\noindent where the hyperparameters of the GP regression model are \\

\begin{ceqn}
\begin{align}
\tag{Eq. 8}
    \Set{\begin{array}{l}
    \mu_*(x) \\
    \Sigma_*(x)
  \end{array}} = \Set{\begin{array}{l}
    \Sigma_0(x, x_{1:k})\Sigma_0(x_{1:k}, x_{1:k})^{-1}(\bar{C_f} - \mu_0(x_{1:k})) + \mu_0(x) \\
    \Sigma_0(x, x) - \Sigma_0(x, x_{1:k})\Sigma_0(x_{1:k}, x_{1:k})^{-1}\Sigma_0(x_{1:k}, x)
  \end{array}}
\end{align}
\end{ceqn}

Once this is done, the BO decides the next condition to test to update the GP regression model at the end, and repeat the process. This choice lies on the acquisition function balancing between the exploitation and the exploration. Most of the time, the choice of the acquisition falls on either the lower confidence bound (LCB), the negative expected improvement (EI), or the negative probability of improvement (PI). In our case study, we have decided to increment a combination of those three function (called Gaussian Process Hedge), by including a probabilistic choice from the past performances of the acquisition functions.

\subsection*{Acquisition function}
\paragraph{} In each step of the BO, the next condition to test $x^*$ depends on the one proposed by the LCB, EI and PI functions. Indeed, the probabilistic choice includes the following steps after initializing a certain gain $(g_i)_{i \leq 3}$ to 0: (i) propose three candidates $\Tilde{x_i}$ from the different acquisition functions, (ii) choose the next conditions $x^*$ by calculating softmax($\eta \times g_i$) ($\eta > 0$), (iii)  update the GP model with ($x^*, y^*$) ($y^* \sim C_f(x^*)$) in order to increase the size of the dataset to calculate the prior knowledge, (iv) update the gains by $g_i = g_i - \mu_0(\Tilde{x_i})$.

\section*{Acknowledgements}
\paragraph{} A.A.F., C.L. and D.Z.D. acknowledge the European Union’s Horizon 2020 research and innovation program for the funding support through the European Research Council (grant agreement 772873, ARTISTIC project). M.D., E.A. and A.A.F. acknowledge the ALISTORE European Research Institute for funding support. A.A.F. acknowledges the Institut Universitaire de France for the support. We acknowledge Dr. Franco Zanotto, Dr. Mehdi Chouchane, and Dr. Teo Lombardo, all from Prof. A. A. Franco’s team for assitance provided for the calculations of conductivities, tortuosity factors, densities and active surfaces.

\section*{Author contributions statement}
\paragraph{} Conceptualization, M.D., and A.A.F.; Methodology, M.D., and A.A.F.; Investigation, M.D., C.L., and D.Z.D.; Software, M.D.; Formal Analysis, M.D.; Writing - Original Draft, M.D.; Writing – Review \& Editing, M.D., C.L., D.Z.D., V.K., E.A., and A.A.F.; Funding Acquisition, E.A., and A.A.F.; Resources, M.D., D.Z.D., and A.A.F.; Supervision, E.A., and A.A.F; Project Administration, A.A.F.

\section*{Code availability}
\paragraph{} Codes and data are available upon reasonable request. They will be made available in a later stage through the computational portal in the \href{https://www.erc-artistic.eu/computational-portal}{ARTISTIC project} website.

\section*{Competing interests}
\paragraph{} The authors declare that they have no known competing interest or personal relationships that could have appeared to influence the work reported in this article.

\bibliographystyle{unsrt}  
\bibliography{ms}

\end{document}


\begin{center}
    \huge \textbf{Machine Learning-Assisted Multi-Objective Optimization of Battery Manufacturing from Synthetic Data Generated by Physics-Based Simulations}
\end{center} 

\vspace{1.5cm}

\maketitle

{$^{1}$ \small Laboratoire de Réactivité et Chimie des Solides (LRCS), Université de Picardie Jules Verne UMR CNRS 7314, Hub de l’Energie, 80039 Amiens, France.}

{$^{2}$ \small Alistore-ERI European Research Institute, CNRS FR 3104, Hub de l’Energie, 80039 Amiens, France.}

{$^{3}$ \small Réseau sur le Stockage Electrochimique de l’Energie (RS2E), FR CNRS 3459, Hub de l’Energie, 80039 Amiens, France.}

{$^{4}$ \small Umicore Group Research \& Development, Olen, Belgium.}

{$^{5}$ \small CIDETEC, Basque Research and Technology Alliance (BRTA), Pº Miramón 196, 20014 Donostia-San Sebastian, Spain.}

{$^{6}$ \small Institut Universitaire de France, 103 Boulevard Saint Michel, 75005 Paris, France.}

{$^{*}$ \small Corresponding author : alejandro.franco@u-picardie.fr (Alejandro A. Franco)}

\hspace{10cm}
\\
\\
\\

\selectlanguage{english}

\newpage
\section{1. Regression plot}
\begin{figure}[!ht] 
	\captionsetup{format=sanslabel}
    \hbox to\hsize{\hss\includegraphics[width=16cm]{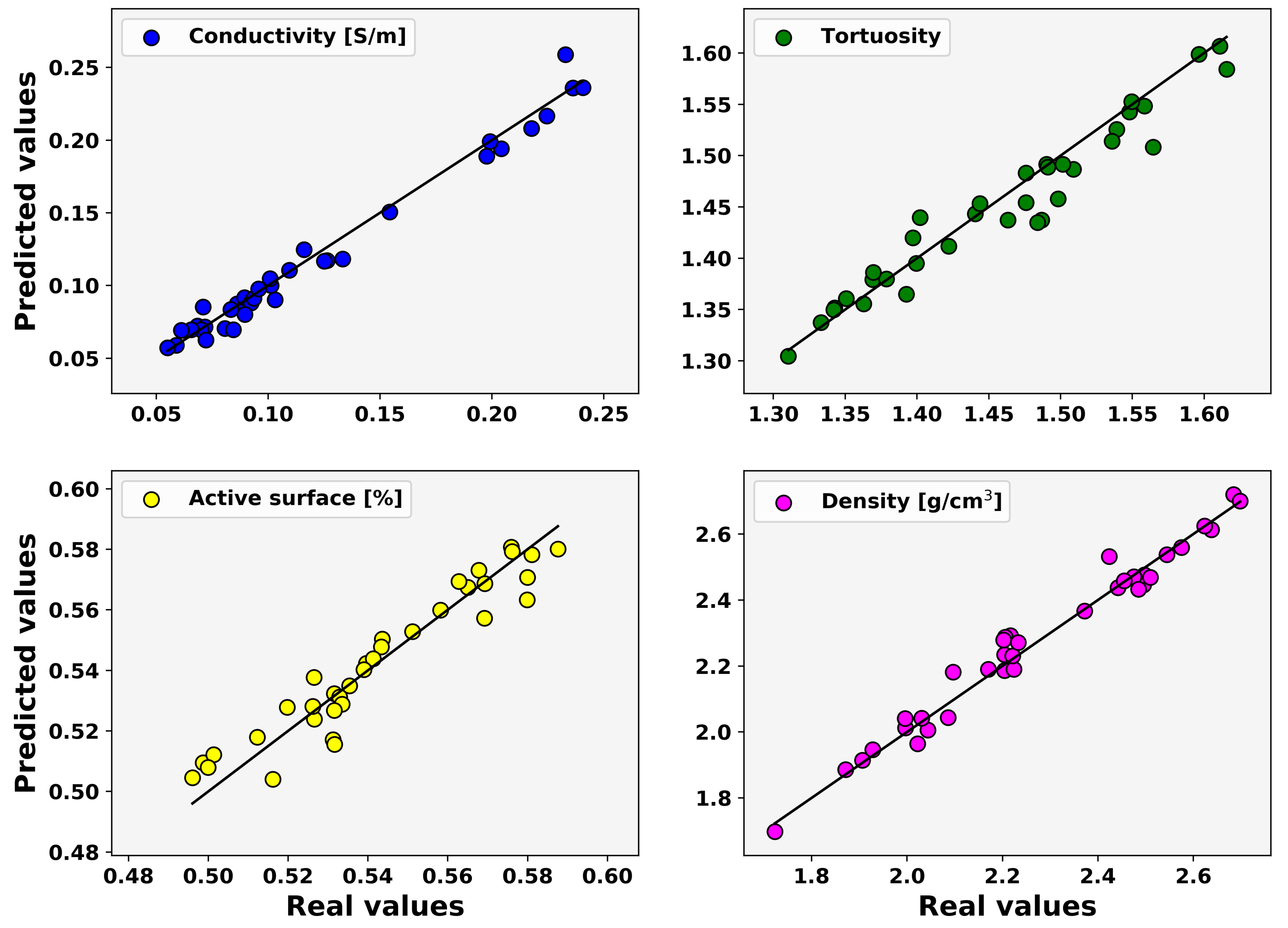}\hss} 
    \caption{ \textbf{Figure S1} : Regression plots displaying the predictive capabilities of the deterministic learning for each electrode property. This evaluates the goodness of fit by comparing the real values from the testing dataset, versus those predicted by the deterministic learnings. This was done for the case of the electronic conductivity (blue), tortuosity factor (green), active surface area (yellow), and density (magenta). }
\end{figure}

\newpage
\section{2. Convergence plot}
\paragraph{} Convergence plot of the minimum value of $C_f$ (cf. equation Eq. 3 in the main manuscript), respect to the number of evaluations n. The decrease is very fast within the first steps and starts to plateauing from step number 100. \\

\begin{figure}[!ht] 
	\captionsetup{format=sanslabel}
    \hbox to\hsize{\hss\includegraphics[width=17cm, height=8.1cm]{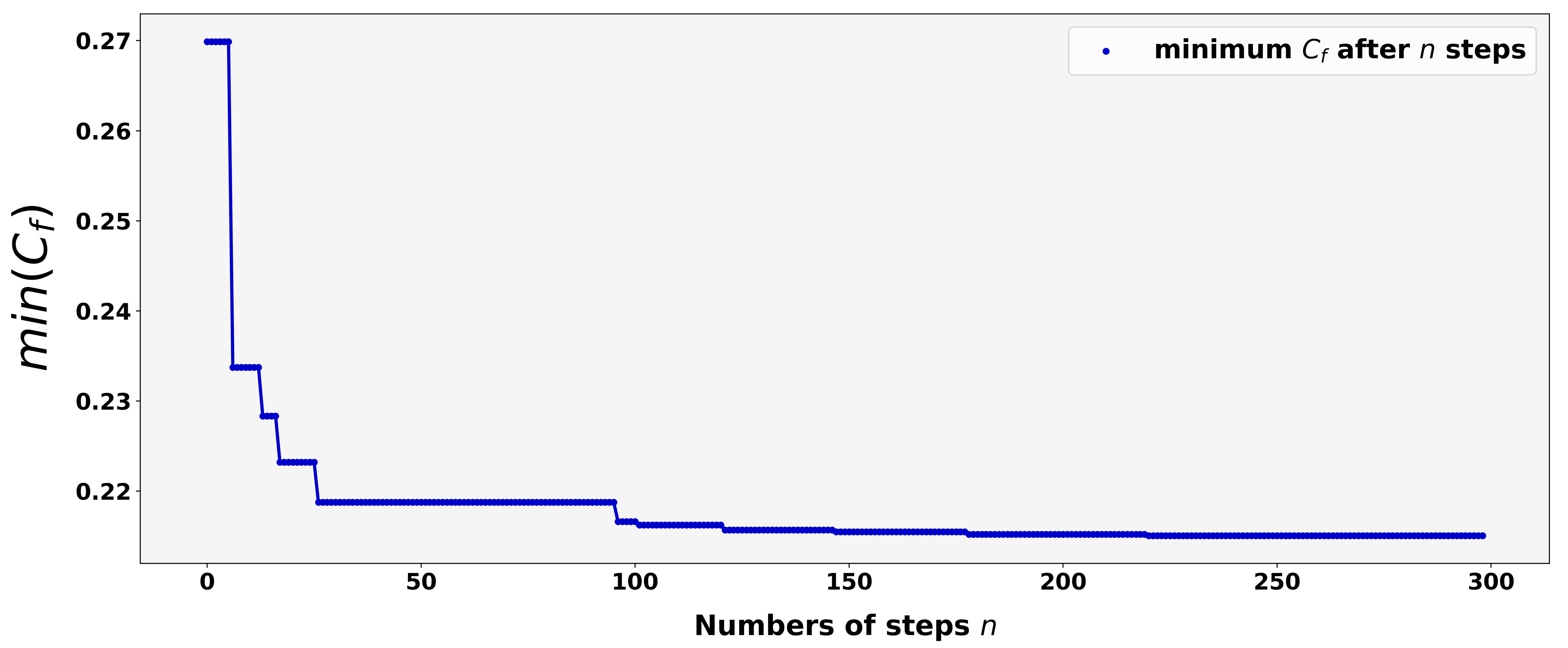}\hss} 
    \caption{ \textbf{Figure S2} : Example of the assessement of output from a KNN model in the case of classificiation and regression for a new data, color-code by a yellow cross. The number of nearest neighbors is set at $k=5$.}
\end{figure}

\newpage
\section{3. Validation metrics related to the training dataset}
\begin{table}[h]
\centering
\small
\setlength\tabcolsep{3pt}
\setlength\extrarowheight{2pt}
\captionsetup{format=sanslabel}

\begin{tabularx}{\textwidth}{ 
  >{\raggedright\arraybackslash}X 
  >{\raggedright\arraybackslash}X
  >{\raggedright\arraybackslash}X
  >{\raggedright\arraybackslash}X}
\midrule
\textsc{Property}   & \textsc{RMSE\%}   & \textsc{$R^2_{score}$} \\ \midrule
Electronic conductivity [S/m]      &  1.91 & 0.966\\
Tortuosity factor      & 6.88 & 0.978\\
Active surface area [\%]      &  1.41 & 0.909\\
Density [g/cm$^3$]      &  1.87 & 0.985\\
    \midrule          
\end{tabularx}
\caption{\textbf{Table S1} : Validation metrics calculated over the training dataset, and associated to the fitting of the electrode properties. The 95\,\% confidence interval (CI95) were estimated with a total of 75 random seeds of training/testing datasets for the uncertainty of the $R^2_{score}$.}
\end{table}

\bibliographystyle{unsrt}